\documentstyle[aps, preprint,eqsecnum, graphicx, verbatim]{revtex}
\tightenlines
\begin{document}
\title{Strong Coupling Fixed Points of Current Interactions and
Disordered Fermions in 2D}
\author{ Andr\'e  LeClair}
\address{Newman Laboratory, Cornell University, Ithaca, NY
14853.}
\address{and}
\address{LPTHE\footnote{Laboratoire associ\'e No. 280 au CNRS},  
4 Place Jussieu, Paris, France.}
\maketitle

\def\betaf{{$\beta$eta~}}

\begin{abstract}

The all-orders \betaf function  is  used to study 
disordered Dirac fermions in $2D$.  The generic strong coupling
fixed `points' of anisotropic current-current interactions
at large distances are actually isotropic 
manifolds corresponding to subalgebras of
the maximal current algebra at short distances.  We argue that
IR 
fixed point theories   are generally
current algebra cosets.  We illustrate this with
the simple example of anisotropic $su(2)$, which is the 
physics of Kosterlitz-Thouless transitions.  
We propose a  phase diagram for the Chalker-Coddington network
model which is in the universality class of the integer
Quantum Hall transition.  One phase is in the universality
class of dense polymers.

\end{abstract}
\vskip 0.2cm
%
%
%
%
\def\oti{{\otimes}}
\def\bra#1{{\langle #1 |  }}
\def\lb{ \left[ }
\def\rb{ \right]  }
\def\tilde{\widetilde}
\def\hat{\widehat}
\def\*{\star}
\def\[{\left[}
\def\]{\right]}
\def\({\left(}		\def\BL{\Bigr(}
\def\){\right)}		\def\BR{\Bigr)}
	\def\BBL{\lb}
	\def\BBR{\rb}
%
%
\def\zb{{\bar{z} }}
\def\zbar{{\bar{z} }}
\def\frac#1#2{{#1 \over #2}}
\def\inv#1{{1 \over #1}}
\def\half{{1 \over 2}}
\def\d{\partial}
\def\der#1{{\partial \over \partial #1}}
\def\dd#1#2{{\partial #1 \over \partial #2}}
\def\vev#1{\langle #1 \rangle}
\def\ket#1{ | #1 \rangle}
\def\rvac{\hbox{$\vert 0\rangle$}}
\def\lvac{\hbox{$\langle 0 \vert $}}
\def\2pi{\hbox{$2\pi i$}}
\def\e#1{{\rm e}^{^{\textstyle #1}}}
\def\grad#1{\,\nabla\!_{{#1}}\,}
\def\dsl{\raise.15ex\hbox{/}\kern-.57em\partial}
\def\Dsl{\,\raise.15ex\hbox{/}\mkern-.13.5mu D}
%
%
\def\th{\theta}		\def\Th{\Theta}
\def\ga{\gamma}		\def\Ga{\Gamma}
\def\be{\beta}
\def\al{\alpha}
\def\ep{\epsilon}
\def\vep{\varepsilon}
\def\la{\lambda}	\def\La{\Lambda}
\def\de{\delta}		\def\De{\Delta}
\def\om{\omega}		\def\Om{\Omega}
\def\sig{\sigma}	\def\Sig{\Sigma}
\def\vphi{\varphi}
%
%
\def\CA{{\cal A}}	\def\CB{{\cal B}}	\def\CC{{\cal C}}
\def\CD{{\cal D}}	\def\CE{{\cal E}}	\def\CF{{\cal F}}
\def\CG{{\cal G}}	\def\CH{{\cal H}}	\def\CI{{\cal J}}
\def\CJ{{\cal J}}	\def\CK{{\cal K}}	\def\CL{{\cal L}}
\def\CM{{\cal M}}	\def\CN{{\cal N}}	\def\CO{{\cal O}}
\def\CP{{\cal P}}	\def\CQ{{\cal Q}}	\def\CR{{\cal R}}
\def\CS{{\cal S}}	\def\CT{{\cal T}}	\def\CU{{\cal U}}
\def\CV{{\cal V}}	\def\CW{{\cal W}}	\def\CX{{\cal X}}
\def\CY{{\cal Y}}	\def\CZ{{\cal Z}}

\def\rvac{\hbox{$\vert 0\rangle$}}
\def\lvac{\hbox{$\langle 0 \vert $}}
\def\comm#1#2{ \BBL\ #1\ ,\ #2 \BBR }
\def\2pi{\hbox{$2\pi i$}}
\def\e#1{{\rm e}^{^{\textstyle #1}}}
\def\grad#1{\,\nabla\!_{{#1}}\,}
\def\dsl{\raise.15ex\hbox{/}\kern-.57em\partial}
\def\Dsl{\,\raise.15ex\hbox{/}\mkern-.13.5mu D}
%
%
%
\font\numbers=cmss12
\font\upright=cmu10 scaled\magstep1
\def\stroke{\vrule height8pt width0.4pt depth-0.1pt}
\def\topfleck{\vrule height8pt width0.5pt depth-5.9pt}
\def\botfleck{\vrule height2pt width0.5pt depth0.1pt}
\def\Zmath{\vcenter{\hbox{\numbers\rlap{\rlap{Z}\kern
0.8pt\topfleck}\kern 2.2pt
                   \rlap Z\kern 6pt\botfleck\kern 1pt}}}
\def\Qmath{\vcenter{\hbox{\upright\rlap{\rlap{Q}\kern
                   3.8pt\stroke}\phantom{Q}}}}
\def\Nmath{\vcenter{\hbox{\upright\rlap{I}\kern 1.7pt N}}}
\def\Cmath{\vcenter{\hbox{\upright\rlap{\rlap{C}\kern
                   3.8pt\stroke}\phantom{C}}}}
\def\Rmath{\vcenter{\hbox{\upright\rlap{I}\kern 1.7pt R}}}
\def\Z{\ifmmode\Zmath\else$\Zmath$\fi}
\def\Q{\ifmmode\Qmath\else$\Qmath$\fi}
\def\N{\ifmmode\Nmath\else$\Nmath$\fi}
\def\C{\ifmmode\Cmath\else$\Cmath$\fi}
\def\R{\ifmmode\Rmath\else$\Rmath$\fi}








\def\beq{\begin{equation}}
\def\eeq{\end{equation}}
\def\cg{{\cal G}}
\def\ch{{\cal H}}

\section{Introduction}  

Two is the critical dimension for Anderson localization, 
namely,   above 2 dimensions states are localized only
above a critical strength of the disorder (impurities)
whereas below 2 dimensions states are localized for
any strength of the disorder\cite{Anderson,Anderson2}. 
For this reason one can expect rich phase structures
in two dimensions.  It is well understood that a complete
understanding of the Quantum Hall transition involves
delocalization\cite{Laughlin}\cite{Halperin}.  In these localization problems, 
understanding the  quantum phase transition in the
conductivity requires finding the right renormalization group
 fixed point
at large distances. This usually occurs at strong coupling,
which is why these problems are considered  difficult. 

The conventional theoretical framework leads to sigma models
on various spaces\cite{Wegner}.   For the quantum Hall transition
the sigma model was worked out by Pruisken\cite{Pruisken}.   
These results are important, but the sigma models have generally 
been 
too difficult to solve. The most recent progress can be found in
\cite{Fendley}.   

Critical theories in 2 dimensions are conformally invariant
and this imposes strong constraints on the theory\cite{BPZ}.    
Early efforts to use conformal field theory methods in the 
study of disordered
Dirac fermions  were made by Bernard\cite{Cargese},  
Mudry et. al. \cite{Mudry}, and Nersesyan et. al.
\cite{Tsvelik}.   Perturbed conformal field theory methods
were also used for disordered statistical mechanical models\cite{Dot}
\cite{Andreas}; generally these are more difficult problems
since the non-random theories are interacting so that one
has to use replicas.        Many of the important problems
here also are driven to
strong coupling
under renormalization group (RG)  flow.  

In this work we describe a general approach to disordered Dirac
fermions at strong coupling which uses the all-orders \betaf functions
proposed in \cite{GLM}.  These models are not integrable, but
one can nevertheless sum up all orders in perturbation theory
for the \betaf function. 
In some insightful work, Chalker and Coddington proposed a simple
network model which is now believed to be in the universality
class of the integer quantum Hall transition\cite{ChalkerC}.  
This network model has been shown to be equivalent to a certain theory
of random Dirac fermions\cite{ChalkerHo},  and this is the model we
analyze using our methods.   This model also appeared in the 
work\cite{Grin}.   A sigma model formulation was proposed in
\cite{Zirnnet}.

This paper is a first  attempt at  understanding  the implications of
these all-orders $\beta$eta functions, and for this reason is
in part conjectural.  
The \betaf functions
 generally do not have non-trivial zeros, but rather have poles.  
We point out that 
\betaf functions with poles are also known to 
occur in supersymmetric gauge
theory\cite{Shifman}. 
In the next section we propose a general scheme for the 
fixed points of marginal  symmetry breaking (anisotropic) 
current-current perturbations for
a current algebra $\cg$  in 2 dimensions based on these 
$\beta$eta functions and certain hypotheses.  
Since there is no small parameter in these theories, one 
generally does not flow to a single fixed point, but rather to
a fixed point manifold in the couplings.  In the case of large
couplings we show  that this
manifold generally corresponds to a sub-current-algebra $\ch$ of   
$\cg$ and argue that the infra-red fixed point is the current
algebra coset $\cg/\ch$.  

The \betaf function shows an interesting
duality, namely $g$ flows to $1/g$ as the length scale varies
from $r$ to $1/r$.  This is a new form of duality since the more
well-known manifestations occur in scale-invariant theories.   
Here, theories at a scale $r$  and coupling $g$ 
are equivalent to a theory
at scale $1/r$ and coupling $1/g$. 

Assuming our hypotheses for interpreting the $\beta$eta functions
are correct, 
they  turn out to predict  a very tight phase
structure; in fact the resulting phase diagram is 
{\it exact} in the sense that the phase boundaries 
are precisely determined. 
We are unaware of any other models where the exact
\betaf function allows one to determine so precisely the phase diagram.       
As an illustration we work out the simplest possible case of 
an anisotropic $su(2)$ perturbation.  This is the subject
of Kosterlitz-Thouless transitions which have previously
only been studied at weak coupling\cite{Kosterlitz}.
The phase diagram is 
unexpectedly rich.   
At small coupling 
one phase exhibits the symmetry restoration recently studied
at one loop\cite{Lin}\cite{Konik}, however it appears this
symmetry does not persist at strong coupling. 
Another phase has a line of 
fixed points corresponding to a free boson at
a certain radius of compactification.

If we extend the physical regime of couplings of the network 
model, the resulting   
disordered Dirac theory again  
has a number of phases. 
Though many of the features are similar to the anisotropic
$su(2)$ case, some features are novel and 
not completely understood.
One  phase ($PSL_g$)  has
a line of fixed points related to $PSL(1|1)$ studied in 
\cite{GLL}, and appears to be the only massless phase. 
It is in the same universality class as dense polymers and is 
also  
 closely related to the disordered XY model\cite{Cardy}. 
A sigma model version of this theory was proposed
by Zirnbauer in connection with the quantum Hall 
transition\cite{Zirnpsl}, based on the work\cite{PSL1}\cite{PSL2}.  
In the physical regime of the network model we show that this
phase is not easily realized since it corresponds to 
imaginary gauge potential.     
The  other  phases  
do not automatically fall into the category of $\cg/ \ch$, 
as explained below.

An unexpected feature of our analysis is that initially positive
couplings which are variances of real disordered potentials can
flow to negative values, which in turn can be viewed as corresponding
to imaginary potentials and thus non-hermitian hamiltonians.
This could signify that our approach breaks down for reasons
we do not yet comprehend.    
If this feature turns out to be sensible, 
it implies  that the studies of delocalization transitions in 
non-hermitian quantum mechanics made by Hatano and Nelson\cite{Nelson} 
may have some bearing on hermitian systems.

\section{General Structure of Fixed Points}

\def\Jb{\bar{J}}

The general class of quantum field theories we consider
are current-current perturbations of a conformally
invariant field theory.   We assume the conformal field
theory possesses left and right conserved currents 
$J^a (z)$ and $\Jb^a (\zbar)$ in the usual way\cite{KZ}\cite{WZW},
where $z, \zbar$ are euclidean light-cone coordinates,
$z=(x+iy)/\sqrt{2}, \zbar = (x-iy)/\sqrt{2}$.  
These currents satisfy the operator product expansion
(OPE):
\beq
\label{IIi}
J^a (z) J^b (0) =  \frac{k}{z^2} \eta^{ab} + \inv{z}
f^{ab}_c \> J^c (0) + ....
\eeq
where $k$ is the level and $\eta^{ab}$ the metric on the
algebra, and similarly for $\Jb$.  
We will refer to this current algebra as $\cg_k$,
where $\cg$ is a Lie algebra or superalgebra 
and the formal action for this theory as $S_{\cg_k}$. 
In the superalgebraic case each current $J^a$ has a 
grade $[a] =0$ or $1$ corresponding to bosonic verses
fermionic.  The metric $\eta^{ab}$ generally
cannot be diagonalized in the superalgebra case. 
The metric and structure constants have the
following properties:
\beq
\label{IIii}
\eta^{ab} = (-)^{[a][b]} \eta^{ba} , 
~~~~~
f^{ab}_c = -(-)^{[a][b]} f^{ba}_c , 
~~~~~
f^{abc} = - (-)^{[b][c]} f^{acb}
\eeq
where $f^{abc} = f^{ab}_i \eta^{ic} $ 
and $\eta_{ab} \eta^{bc} = \delta^c_a$.  

\def\ab{\bar{a}}
\def\bb{\bar{b}}
\def\dx{\frac{d^2 x}{2\pi} }

The theory $S_{\cg_k}$ can be perturbed by marginal
operators built out of left-right current bilinears:
\beq
\label{IIiii}
S =  S_{\cg_k} + \int \dx ~ \sum_A g_A \> \CO^A , 
~~~~~~~~~~~~  \CO^A = \sum_{a, \bar{a}} d^A_{a\ab} J^a \Jb^{\ab} 
\eeq
The simplest example is a single coupling $g$ with 
$d_{a\ab}= \eta_{a\ab} $ defining the quadratic Casimir of $\cg$.  
For $su(N)$ at level $1$, this is the non-abelian Thirring
model, or equivalently the chiral Gross-Neveu
model.   We are mainly interested in more general
situations where the tensors $d^A_{ab}$  break the symmetry
$\cg$, i.e. are anisotropic.  

\def\betaf{$\beta$eta~}

In \cite{GLM} an all-orders   \betaf function in a certain
 prescription was proposed and we now summarize
this result.  
The theory is not renormalizable for any choice of 
$d^A$.   The three conditions ensuring renormalizability
are:
\begin{eqnarray}
\label{IIiv}
(-)^{[b][c]} d^A_{ab} d^B_{cd} f^{ac}_i f^{bd}_j 
 &=& C^{AB}_C \> d^C_{ij} 
\\ 
\eta^{ij} d^A_{ai} d^B_{bj} &=& D^{AB}_C \> d^C_{ba} 
\\
d^A_{ij} f^{ja}_{k}f^{ik}_b &=& R^A_B \> \eta^{ac} d^B_{cb} 
\end{eqnarray}
The first condition is equivalent to closure of the
operator algebra of $\CO^A$:
\beq
\label{2.5}
\CO^A (z, \zbar) \CO^B (0) \sim \inv{z\zbar} 
\sum_C C^{AB}_C \> \CO^C (0) 
\eeq
and guarantees one-loop renormalizability.  
The other two conditions are necessary at 
two loops and higher.  The structure constants 
$D, R$ are also related to an OPE.  Define the 
operator $T^A$ built out of left-moving currents only
 and $d^A$:
\beq
\label{IIvi} 
T^A (z)  = d^A_{ab} J^a (z)  J^b (z)  
\eeq
The operator $T^A$ is a kind of stress tensor;
in the isotropic case it is the affine-Sugawara stress tensor
up to a normalization.  One finds
\beq
\label{2.7}
T^A (z) \CO^B (0) \sim \inv{z^2} 
\( 2k D^{AB}_C +  R^A_D D^{BD}_C \) \CO^C (0) 
\eeq
Given a particular theory, the above equation is
an efficient way to compute $R,D$;  the $D$ term
is distinguished from the $RD$ term by being 
proportional to $k$.  
The conditions required for $T^A$ to satisfy the
OPE of a consistent stress-tensor, the so-called master
equation studied extensively in \cite{Halpern}, 
involves the same objects $C, D, R$, but 
appears to be a stronger than our renormalizability
conditions.     

The renormalization group (RG) structure constants 
have the following properties:
\beq
\label{IIxib}
D^{AB}_C = D^{BA}_C ,  ~~~~~~ 
C^{AB}_C = C^{BA}_C, ~~~~~ 
D^{AC}_D D^{DB}_E = D^{AB}_D D^{DC}_E 
\eeq
which can be proven from the defining relations
or from the OPE's.

\def\gh{{\hat{g}}}

Let us arrange the couplings into a row vector
$g = (g_1, g_2, ...)$.  Since the higher loop
expansion is in $kg$, let us define
\beq
\label{2.8}
\gh = kg/2 
\eeq
Let $D(\gh)$ be the matrix of couplings
\beq
\label{2.9} 
D(\gh)^A_B = \sum_C  D^{AC}_B \> \gh_C  
\eeq
Given two row vectors $v^{1,2}$ we define a new
row vector from $C$:
\beq
\label{2.10}
C(v^1, v^2 )_A = \sum_{B,C}  v^1_B v^2_C  ~ C^{BC}_A  
\eeq
Finally, define 
\beq
\label{2.11}
\gh' = \gh \inv{1-D(\gh)^2 } 
\eeq
Then the \betaf function
can be expressed as
\beq
\label{2.12}
\beta_\gh  = \frac{ d\gh}{d\log r} = \frac{2}{k}  
\(  
-\inv{2} C(\gh' , \gh') (1+ D^2) + C(\gh' D, \gh'D )  D 
- \gh' D R D \) 
\eeq
where $r$ is a length scale and 
$D = D(\gh)$.  The flow to the IR corresponds to increasing $r$.   

\def\ct{\tilde{C}}

For the purpose of studying fixed points, it will be
convenient to define 
\beq
\label{2.13} 
\ct^{AB}_C = R^A_D D^{BD}_C 
\eeq
Note that $\ct$ appears directly in the OPE (\ref{2.7}) 
and the \betaf function so there is no need to 
compute $R$ by itself.  One can show 
\beq
\label{2.14} 
\ct (v^1 , v^2 D(\gh) ) = \ct (v^1 , v^2 ) D(\gh )
\eeq
using (\ref{IIxib}). 
The \betaf function can then be arranged into the form:
\begin{eqnarray}
\label{2.15}
\beta_\gh = \inv{k} && 
\Bigl[ 
- C(\gh'  , \gh' ) (1-D)^2 + \ct(\gh_+ , \gh_- ) 
- \ct (\gh_- , \gh_+ ) 
\\
~~~~~~~~~~~&&
- \( (C+\ct )(\gh_- , \gh_+ )\) D  
- \( (C + \ct ) (\gh_+ , \gh_- ) \) D   
\Bigr] 
\end{eqnarray}
where again $D = D(\gh)$ and we have defined  
\beq
\label{2.16}
\gh_\pm = \gh \inv{1\pm D(\gh )} 
\eeq

We now study the possible fixed points based on the
above \betaf function.  In the specific examples we
have studied $\beta_{\gh}$  has no non-trivial 
zeros for any finite values
of $g$. 
The only zeros of $\beta_{\gh}$ in the two examples
below  are trivial in the sense that they correspond to $g=0$ 
or  exactly marginal directions,  such as $u(1)$ directions.
We believe this is the generic situation but
cannot rule out other examples with non-trivial zeros.

The \betaf function generally has poles as can be seen
from (\ref{2.11}).  These arise from summing geometrical
series in perturbation theory.  Since these series do not
converge for $\gh >1$ one should question their validity 
for large $\gh$.  This being the first attempt to make sense
of these \betaf functions,  in the sequel we will simply assume the above
\betaf function is valid for all $g$, which amounts to 
an analytic continuation of the perturbation series.  
As we will see, the resulting RG flows can be interpreted at
large $g$.

Assuming the \betaf has no non-trivial zeros, 
then this  implies that we must look for fixed point
manifolds.  Suppose all couplings $\gh$ have large 
absolute value.  Since $\gh' \sim 1/\gh $, 
$\gh_\pm \sim 1$ and $D(\gh) \sim \gh$, the
\betaf function is dominated by the $C+\ct$ terms
which are of order $\gh$.   For large $g$ we therefore expect 
the possibility of flowing to manifolds in the
coupling constant space satisfying 
\beq
\label{2.17b}
(C + \ct ) (\gh_- , \gh_+ ) 
= (C + \ct ) (\gh_+ , \gh_- ) = 0 
\eeq
In the examples below the solutions to the above equations
are the same as for 
\beq
\label{2.17}
  C(g, g) + \ct (g, g) = 0 
\eeq

The above argument is not strong enough to conclude that the
{\it only}  asymptotic flows  are  solutions to (\ref{2.17}).
Rather, we expect that solutions to (\ref{2.17}) 
represent a generic class of fixed points.  This 
is strongly supported by the numerical analysis of
the exact \betaf functions in the examples below; it is
found that   when
couplings flow to infinity, 
one indeed generally flows to solutions of the equation
(\ref{2.17}).    

Solutions to the above fixed point manifold
equation are generally couplings that correspond to invariant
subalgebras of $\cg$.  Let $Q^a$ denote 
generators of $\cg$:
\beq
\label{2.18} 
[Q^a, Q^b \} = f^{ab}_c Q^c  
\eeq
In the conformal field theory these are realized
as conserved charges:
\beq
\label{2.19}
Q^a = \oint \frac{dz} {2\pi i } \> J^a (z) 
+ \oint \frac{d\zbar}{2\pi i } \> \Jb^a (\zbar) 
\eeq
Suppose the $d^A_{ab}$ define invariants of $\cg$: 
\beq
\label{2.20}
\[ Q^a , d^A_{bc} Q^b Q^c \] = 0
\eeq
Then this implies 
\beq
\label{2.21}
d^A_{ib} f^{ai}_c = (-)^{[a][c]} d^A_{ci} f^{ai}_b 
\eeq
Using this in the renormalizability conditions 
(\ref{IIiv}) 
one finds that this implies $C + \ct =0$.

Interesting possibilities for non-trivial fixed points
thus arise when the $d^A$ couple commuting subalgebras
of $\cg$.  Let 
\beq
\label{2.22}
\CO^A  =  \eta^A_{ab} J^a_A \Jb^b_A 
\eeq
where $J_A$ are the currents for a subalgebra $\ch^A$ 
of $\cg$, $\eta^A$ defines the Casimir for $\ch^A$, 
and these subalgebras commute for $A \neq B$. 
The currents $J_A$ satisfy a current algebra 
$\ch^A_{k_A} $, where the level
$k_A$ is generally not equal to 
$k$.  This is sometimes referred to as a conformal 
embedding.   Only the currents $J_A$ matter in
(\ref{2.20}) and the fixed point manifold condition
$C+\ct =0$ is satisfied.   

Let us further suppose that each $\ch^A$ has a
unique quadratic Casimir so that 
\beq
\label{2.23} 
\eta^A_{ij} f^{jc}_k f^{ik}_d  = C^A_{adj} \delta^c_d
\eeq
where $C^A_{adj}$ is the quadratic Casimir in the 
adjoint representation of $\ch^A$.  Not all algebras
have this property; in particular the superalgebras
with indecomposable representations which we will
encounter below do not.  
Assuming (\ref{2.23}), one finds that 
\beq
\label{symmetric}
C^{AA}_A = -C^A_{adj} , ~~~~~ R^A_B = C^A_{adj} \delta^A_B, 
~~~~~ D^{AA}_A = 1
\eeq
The \betaf functions then obviously decouple and take 
the simple form:
\beq
\label{2.24}
\beta_{\gh_A}  =  \frac{C^A_{adj}}{k} \frac{ \gh^2_A}
{ (1+ \gh_A )^2 } 
\eeq

The \betaf function (\ref{2.24}) has some interesting
properties.   Let us focus on a single coupling 
$\gh_A = \gh$.  Define a dual coupling 
\beq
\label{dual}
g^*  =  1/ \gh 
\eeq
The \betaf function satisfies 
\beq
\label{2.25}
\beta^* (g^*) = - \beta(\gh \to g^*) 
\eeq
This implies that 
\beq
\label{2.26}
\gh (r) =  \inv{\gh(1/r)}
\eeq
This behavior can be seen from the solution:
\beq
\label{2.27}
(\gh - 1/\gh )/2 + \log \gh =  \frac{C_{adj}}{2k} \log r/r_0 
\eeq
where $r_0$ is the scale where $\gh = 1$.  
As shown below, the anisotropic case also exhibits this
kind of duality.

There are four cases to consider.  First suppose
$C_{adj}/k  > 0$.  Then $\gh$ always increases.
From the duality (\ref{2.26})  one sees that 
$\gh >0$ at short distances (small $r$) flows 
to $\gh = \infty$ at large $r$.  On the other hand any 
$\gh <0$ flows to $0^-$.  Thus $\gh>0$ is marginally
relevant and $\gh<0 $ marginally irrelevant.  This
confirms the expectations based on one-loop.  
Note that the double pole in the \betaf function 
at $\gh = -1$ has no physical effect whatsoever:
near the pole $\gh$ grows faster as $r^{C_{adj}/2k}$
rather than  $\log r$ but simply flows through the pole. 
When $C_{adj}/k <0$ everything is reversed:
$\gh >0$ flows to $0^+$ and $\gh <0$ flows to 
$-\infty$.   Thus in this case of isotropic couplings, 
the \betaf function appears to be physically
sensible  for all $\gh$, and this supports the validity
of the analytic continuation of the \betaf function.    

We thus propose the following generic scheme for
non-trivial infrared (IR)  fixed points that can occur when
couplings flow to infinity.  
For every marginally relevant $g_A$, $g_A$ flows
to either $\pm \infty$.  The degrees
of freedom coupled by $g_A$ are thus massive 
and should decouple in the flow to the IR.   
This picture was used in the work \cite{SQHE} on
the spin quantum Hall effect.   For example consider
an $su(3)_1$ current algebra perturbed by the marginal operator
that isotropically only couples an $su(2)_1$ sub-current algebra
with coupling $g$.
The $su(3)_1$ current algebra can be bosonized in terms of 
two free bosons $\phi_1,  \phi_2$ and has $c=2$.  
The $su(2)_1$ current algebra can be bosonized in terms of 
a single scalar field $\phi$ which is a linear combination of
$\phi_1 , \phi_2$, and has $c=1$.  
 As $g$ goes to $\infty$ under RG flow,  correlation functions
of $\phi$ tend to zero.   Thus the field $\phi$ decouples
and the IR fixed point is $su(3)_1/su(2)_1$.   
Generally, the fixed point is thus 
\beq
\label{fixed}
{\rm IR ~ fixed~ point} = ~~ 
\frac{\cg_k}{\ch_{k}}  ,  ~~~~~~
\ch_{k} = \otimes_{A} \ch^A_{k_A} 
\eeq 
where $\otimes_A$ is only over marginally relevant $g_A$. 
The IR theory is non-trivial only if $\ch \neq \cg$. 
This situation corresponds to a massless phase and 
$\cg_k /\ch_{k} $ are the massless degrees of freedom.  
When $\cg_k / \ch_k$ is empty, e.g. when $\cg = \ch$,
 then this is a massive phase.
In localization problems a massive phase corresponds to all
states being localized.  
Coset conformal field theories were studied in generality 
in \cite{GKO}.
The Virasoro central charge\cite{BPZ} at the fixed point
is $c^{\cg} - c^{\ch}$, and the  conformal scaling 
dimension $\Delta$ of an operator at the fixed point is
\beq
\label{dimension}
\Delta^{\cg/\ch} = \Delta^\cg - \Delta^\ch 
\eeq
\
An interesting open question is whether
other  models can flow to the more  general solutions of the master
equation\cite{Halpern} which do not correspond to cosets.

Due to the existence of poles in the \betaf function, 
not all flows that we find in the examples below fall
neatly into the above scheme. In some cases  
below the RG flows are attracted to these poles.  
In most of 
 these cases, as one approaches the pole, some of the
other couplings flow off to infinity in such a way 
that suggests one can still perform a 
coset.

\section{Anisotropic $su(2)$ } 

In this section we illustrate the scheme of the last section
in the simplest possible example of an anisotropic $su(2)$.  
This is the subject of Kosterlitz-Thouless  flows which
have previously only been studied at weak coupling\cite{Kosterlitz}.
We  originally 
worked out this example to check of the validity of
the exact $\beta$eta functions.  The phase diagram is 
unexpectedly rich.

\def\barray{ \begin{eqnarray} }
\def\earray{ \end{eqnarray} }

We normalize the currents as follows:
\beq
\label{4.1} 
J(z) J(0) \sim \frac{k}{2} \inv{z^2} , ~~~~~~~
J(z) J^\pm (0) \sim \pm \inv{z} J^\pm (0) 
,~~~~~
J^+ (z) J^-(0) \sim \frac{k}{2} \inv{z^2} + \inv{z} J(0) 
\eeq
and consider the action
\beq
\label{4.2}
S  =  S_{su(2)_k}  +  \int \dx \(  g_1 (J^+ \Jb^- + J^- \Jb^+ ) 
+ g_2 J \Jb  \) 
\eeq
A simple computation using (\ref{2.5}, \ref{2.7})  gives the RG data:
\barray
\label{rgdata}
C^{12}_1 &=& C^{21}_1 = -1, ~~~~~C^{11}_2 = -2
\\
\ct^{11}_1 &=& \ct^{21}_1 = 1, ~~~~~\ct^{12}_2 = 2
\\  
D^{11}_1 &=& D^{22}_2 = 1/2
\earray
The matrix D(g) is diagonal:
\beq
\label{dgmatrix}
D(g)  =  \left(\matrix{g_1 /2  & 0 \cr 0 & g_2 /2 \cr} \right)   
\eeq
The resulting \betaf functions are 
\barray
\label{4.3}
\beta_{g_1} &=&  \frac{ g_1 (g_2 - g_1^2 k/4 )}
                   {( 1-k^2 g_1^2/16 )(1+kg_2/4 ) }
\\
\beta_{g_2} &=&  \frac {  g_1^2 (1-kg_2/4)^2  }
                         { (1-k^2 g_1^2/16 )^2 } 
\earray

There is again an interesting duality in these \betaf functions.
Define the dual couplings
\beq
\label{dualg}
g_1^* = \frac{16}{k^2 g_1}  , ~~~~~~~
g_2^* = \frac{16}{k^2 g_2} 
\eeq
Then the \betaf function satisfies
\beq
\label{andual}
\beta^* (g^*) = - \beta(g\to g^*)
\eeq

The above \betaf functions  predict a tight but unexpectedly
rich phase diagram.  The duality
(\ref{andual}) explains some features of the diagram,
namely, the self-dual lines $g = g^*$ are phase boundaries
or lines of attraction.   
The solutions to $C+\ct = 0$ are 
(i) $g_1 = g_2$ , which corresponds to the subalgebra
 $\ch_k  = su(2)_k$ 
and (ii) $g_1 = 0$ corresponding to $\ch = u(1)$. 

The structure of the phase diagram is determined in part  by the
behavior near the poles at $g_1, g_2 =\pm 4/k$.  
Consider the pole at $(g_1, g_2 ) = (4,4)/k$ and
let $g_{1,2} = 4(1+ \ep_{1,2} )/k$.  Near $\ep = 0$ the
behavior is 
\beq
\label{4.4}
\beta_{\ep_1} \approx 8-\frac{4\ep_2}{\ep_1} , ~~~~~~
\beta_{\ep_2} \approx  \frac{4\ep_2^2}{\ep_1^2} 
\eeq
Around this pole, this leads to the behavior shown in figure 1.   
In this figure,  heavy lines correspond to phase boundaries. 
The region, or phase, $A$ is attracted to the line $g_1 = g_2 $,
since this line is stable in region A. 
Beyond $(g_1, g_2 ) = (4,4)/k$ the line $g_1 = g_2$ becomes
unstable.  
Since the line becomes
unstable beyond the pole, one has to reach the line {\it exactly}
in the region $A$ before flowing off to infinity along it; otherwise
one can flow elsewhere (see below).  If one is on the line, then
one flows to infinity and the IR fixed point is the empty coset
$su(2)_k/su(2)_k$.

In region $D$, $g_2$ flows to $\infty$ whereas $g_1$ flows
to a constant, but after a finite scale transformation. 
Because one coupling is blowing up, the flow cannot be continued
to larger scales numerically.       
Since the ratio $g_1/g_2$ flows to 
zero, one possible interpretation of
this flow is that   $g_1$  is effectively  zero
and $g_2 = \infty$.  The subalgebra coupled by $g_2$ is
$\ch = u(1)$, and the IR fixed point would thus be  
$su(2)_k/u(1)$.
This is a well-known conformal
embedding\cite{FatZam} and corresponds to the critical theory
of $Z_k$ parafermions with central charge $c=2(k-1)/(k+2)$,
where $k=2$ is the Ising model.   
However we emphasize that since the RG flow cannot be continued
to arbitrarily large length scales,  this is not a true fixed point.

\def\bh{\hat{\beta}}

In phase $E$, $g_1$ grows to infinity and $g_2$ to a 
non-universal value 
$0<g_2 < 4$.  Here the flow can be continued to arbitarily large
scales.  For $k=1$ we interpret this as a sine-Gordon
phase.  At $k=1$ we can bosonize the currents
\beq
\label{4.5}
J^\pm = \inv{\sqrt{2}} e^{\pm i \sqrt{2} \vphi } , 
~~~~~ J = \frac{i}{\sqrt{2}} \d_z \vphi
\eeq
where $\vphi (z) $ is the  $z$-dependent  part of a free massless
scalar field $\phi$.  
Viewing the $g_2$ coupling as a perturbation of the
kinetic term and rescaling the field $\phi$ one
obtains the sine-Gordon action
\beq
\label{4.6}
S = \inv{4\pi} \int d^2 x \[ \inv{2} (\d\phi)^2 
+ g_1 \> \cos (\hat{\beta} \phi ) \]
\eeq

The region A is also sine-Gordon like
at small coupling.       
The operator $\cos \hat{\beta}$ has scaling dimension 
$\Gamma = \bh^2$.   
To relate $\bh$ to $g_2$, consider the \betaf function when 
$g_1$ is small.   Then  
$\beta_{g_1} = g_1 g_2 /(1+g_2 /4)$.  Since this is proportional
to $g_1$, we can identify the dimension of the coupling
as ${\rm dim} (g_1) = g_2/(1+g_2/4) = 2-\Gamma$.  
Thus:
\beq
\label{scale}
\frac{\bh^2}{2}  =  \frac{1-g_2/4}{1+ g_2/4}, ~~~~~~~{\rm for~} 
g_1 \approx 0
\eeq
In general,  we expect $\bh$ to be a function of both $g_1,g_2$.
Note that for small $g_1$, 
values of  $\gh$  between $0$ and
$4$ correspond to 
 $0<\bh^2 < 2$ which is the expected regime for the sine-Gordon
model. (The conventional sine-Gordon coupling $\beta$  is 
 $ \beta = \bh \sqrt{4\pi} $.)  
For $k>1$ the appropriate generalization is the fractional
super sine-Gordon model\cite{FSSG}.

In the $G$ phase $g_1$ flows to zero and $g_2$ to 
a fixed value on the line $-4<kg_2 <0$.
For example, as $r\to \infty$ one finds
\beq
(g_1 , g_2 ) = (2, -3) \longrightarrow (0, -2.337693444..)
\eeq
when $k=1$. 
This line of
fixed points corresponds to a free boson at some radius
of compactification determined by the value of $g_2$ on
the fixed line.      

In region I one is attracted to the isotropic line and then
flows toward the poles at $(k g_1,k g2) = (4,-4)$.   

\begin{figure}[htb]
\hspace{18mm}
\includegraphics[width=13cm]{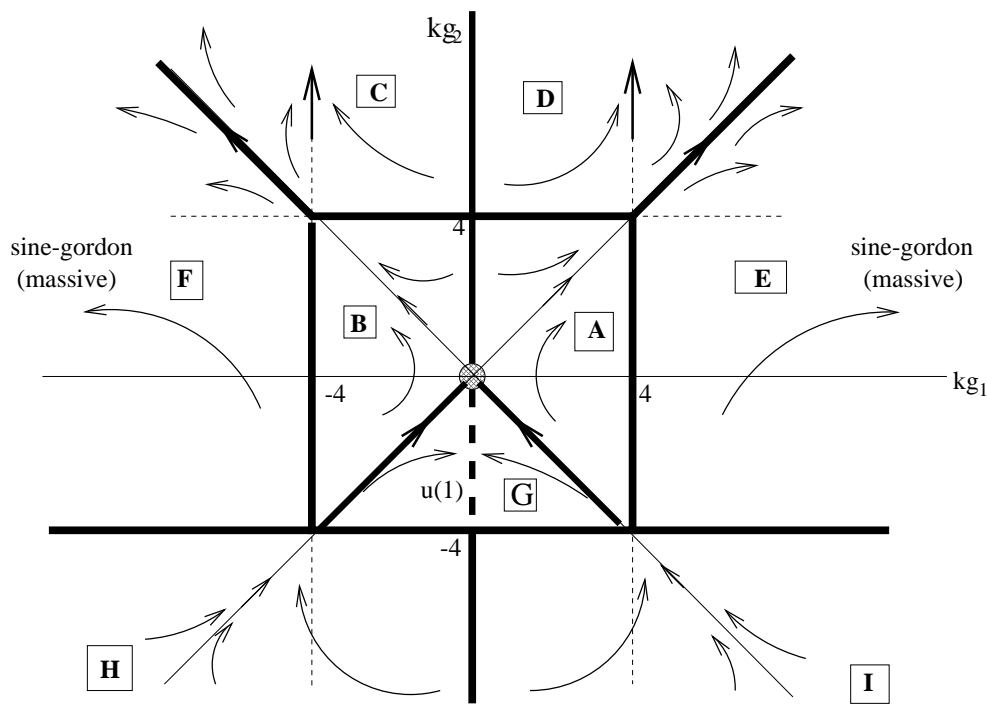}
\caption{Phase Diagram for anisotropic $su(2)$.  Phase boundaries
are solid heavy lines.  Heavy dashed lines denote lines of fixed points.
Shaded circles are fixed points.}
\label{Figure1}
\end{figure}

The flows toward the poles in region A and I is rather delicate,
and the outcome cannot be studied numerically by solving the differential
equations because of the singularity. 
As stated above, one possibility is to continue to flow off to
infinity along the isotropic lines once one reaches the poles.  
However this appears to be inconsistent with the duality (\ref{andual}). 
The duality implies that if a flow passes through the self-dual
point  $g=g^*$,  then $g$ and $g^*$ are on the same RG trajectory, 
where  $g$ flows to $g^*$ as $r$ goes to $1/r$.  
We can use this to extend the flows through the poles\footnote{An 
earlier version of this paper did not consider this possibility,
which will be further reported on in \cite{BL2}.}. 
Namely,
the region in A below the isotropic line flows to the region D,
whereas the region above the line flows to region E.   
Similarly, the region above the isotropic line in I flows to region G,
whereas the region below the line flows to A then to D.  
This implies that the symmetry restoration seen in region A
is destroyed once the flow reaches beyond the pole.

The remaining phases are a mirror image of the above.  
Though the line $g_1= -g_2$ does not appear as a solution
to $C + \ct =0$, and it would naively break the $su(2)$
symmetry, it turns out that it does possess an $su(2)$
symmetry.  There is an automorphism of $su(2)$ 
$J^\pm \to - J^\pm$ that preserves the algebra.  We
can perform this automorphism on the left currents 
{\it only}.  This takes $\CO^1 \to - \CO^1$, i.e. flips
the sign of $g_1$.  The phase diagram then has symmetry
with respect to reflections about the $g_2$ axis.  
Thus the $C,F,B, I$ phases are similar to the 
$D,E,A, H$ phases.

A check of the above \betaf function is based on comparison
with the sine-Gordon \betaf function which is known to
two loops.  For a recent discussion, see \cite{Balog}.
These perturbative computations were performed around
$\bh = \sqrt{2}$ and $g_1$ equal to zero.  Define new
couplings $\delta, \alpha$ as
\beq
\label{SG1}
1 + \delta = \frac{\bh^2}{2}  =  \frac{1-g_2/4}{1+ g_2/4},
~~~~~~ \alpha = 4g_1
\eeq
Then the \betaf function (\ref{4.3})  implies the following
\betaf functions for $\delta, \alpha$ to 5 loops:
\barray
\nonumber
\beta_\delta &=&  -\inv{32} \alpha^2 - \inv{16} \alpha^2 \delta
- \inv{4096} \alpha^4 - \inv{32} \alpha^2 \delta^2 - \inv{2048} 
\alpha^4 \delta 
\\
\label{SG2} 
\beta_\alpha &=& -2\alpha \delta - \inv{64} \alpha^3 - \inv{64}
\alpha^3 \delta - \inv{16384} \alpha^5 
\earray
It is known that with more than one coupling, not all
2 loop contributions are prescription independent.
Define new couplings to second order as
\beq
\label{SG3}
\delta' = \delta - \inv{32} \alpha^2 , ~~~~~~~~\alpha' = \alpha
\eeq
Then the \betaf function (\ref{SG2}) to two loops leads to
\barray
\nonumber
\beta_{\delta'} &=& -\inv{32} \alpha'^2 + \inv{16} \alpha'^2 \delta'
\\
\label{SG4}
\beta_{\alpha'} &=& -2\alpha'\delta' - \frac{5}{64} \alpha'^3
\earray
This agrees with the result presented in \cite{Amit}\cite{Balog}.

\section{The Network Model}  

\subsection{\betaf function }

The network model can be mapped onto a model of
disordered Dirac fermions\cite{ChalkerHo}.  The two-component
hamiltonian in the two spacial dimensions $x,y$  is 
\beq
\label{3.1}
H = \inv{\sqrt{2}} \( -i \d_x - A_x \)\sigma_x 
+ \inv{\sqrt{2}} \( -i \d_y - A_y \) \sigma_y + V + M \sigma_z 
\eeq
where $\sigma$ are Pauli matrices and $A(x,y), V(x,y),
M(x,y)$ are real disordered potentials.  Disorder in $A,V$ and $M$
respectively 
corresponds to randomness in the individual link phases,
the total Aharanov-Bohm phase per plaquette,  and the tunneling
at the nodes.  
This model was also considered in the work \cite{Grin}.  

\def\psib{\overline{\psi}}

Introducing 2-component Dirac fermions 
\beq
\label{3.2}
\Psi = \left( \matrix{ \psi_+ \cr \psib_+ \cr } \right) 
, ~~~~~~~
\Psi^\star = (\psib_- , \psi_- ) 
\eeq
one needs to study the action 
\barray
S &=&  i \int \dx ~ \Psi^\star H \Psi 
\\
\label{3.3}
&=&  \int \dx  
\Bigl[  \psib_- (\d_z - i A_z ) \psib_+ + \psi_- (\d_\zbar 
-i A_\zbar ) \psi_+  
\\  
&& ~~~~~~~~~~~~~-i V (\psib_- \psi_+  + \psi_- \psib_+ ) 
    ~ -i M ( \psib_- \psi_+ - \psi_- \psib_+ ) 
\Bigr]
\earray
where $z=(x+iy)/\sqrt{2} , \zbar = (x-iy)/\sqrt{2} $
and $A_z = (A_x - i A_y)/\sqrt{2} , A_\zbar =
(A_x + i A_y )/\sqrt{2} $.  

We take the potentials to have the following gaussian
distributions:
\barray
\label{3.4} 
P[V] &=& \exp \( -\inv{2 g_v} \int \dx V^2  \) 
\\
P[M] &=& \exp \( -\inv{2 g_m} \int \dx M^2  \)
\\
P[A] &=& \exp \( -\inv{ g_a} \int \dx A_z A_\zbar  \)
\earray
The couplings $g_v, g_m , g_a$ are positive variances of
the potentials. If $A\to i A$, then $g_a \to - g_a$,
and similarly for $M,V$.  Thus negative $g$'s can be
interpreted as corresponding  to imaginary potentials.  

\def\betab{\overline{\beta}}

Since we are dealing with a free theory, we can use
the supersymmetric method for disorder averaging\cite{Efetov}.  
In the present context this method was studied in
\cite{Cargese}\cite{Mudry}.  Introducing bosonic ghost
partners $\beta_\pm , \betab_\pm$ of the fermions and performing
the gaussian integrals one obtains:
\beq
\label{eff}
S_{\rm eff}  = S_{\rm free} 
+ \int \dx  \(  g_v \CO^v  + g_m \CO^m + g_a \CO^a \)
\eeq
$S_{\rm free}$ is a free $c=0$ conformal field theory 
with the action
\beq
\label{3.5}
S_{\rm free}  = \int \dx  \( 
\psib_- \d_z \psib_+ + \psi_- \d_\zbar \psi_+ 
+ \betab_- \d_z \betab_+ + \beta_- \d_\zbar \beta_+ \)
\eeq
A treatment of the $c=-1$ ghost system can be found in
\cite{FMS}.  

\def\Sh{\hat{S}}

The maximal conserved currents of $S_{\rm free}$ are the 
$8$ possible bilinears in the fermions and ghosts:
\beq
\label{3.6}
H = \psi_+ \psi_- , ~~~~~ J = \beta_+ \beta_- , ~~~~~ 
J_\pm = \beta_\mp^2 ,~~~~~
S_\pm = \pm \psi_\pm \beta_\mp , 
~~~~~\Sh_\pm = \psi_\mp \beta_\mp 
\eeq   
Using the OPE's:
\barray
\psi_+ (z) \psi_- (0) &\sim& \psi_- (z) \psi_+ (0) \sim 1/z 
\\ \nonumber
\beta_+ (z) \beta_- (0) &\sim& - \beta_- (z) \beta_+ (0) 
\sim 1/z 
\label{3.7}
\earray
one finds that the currents satisfy the $osp(2|2)_{k=1}$
current algebra:
\begin{eqnarray}
\nonumber
J(z) J(0) &\sim& - \frac{k}{z^2} ,  
~~~~~~~~~~~~~~~H(z) H(0) \sim \frac{k}{z^2}
\\ \nonumber
J(z) J_\pm (0) &\sim&  \pm \frac{2}{z} ~ J_\pm
, ~~~~~~~~~~
J_+(z) J_- (0) \sim \frac{2k}{z^2} - \frac{4}{z} J
\\ \nonumber
J(z) S_\pm (0) &\sim&  \pm \inv{z} S_\pm   ,  
~~~~~~~~~~ J(z) \Sh_\pm (0) \sim
\pm \inv{z} \Sh_\pm
\\ \nonumber
H(z) S_\pm (0) &\sim&  \pm \inv{z} S_\pm  , ~~~~~~~~~~
H(z) \Sh_\pm (0) \sim \mp \inv{z} \Sh_\pm
\\ \label{3.8}
J_\pm (z) S_\mp (0) &\sim& \frac{2}{z} \Sh_\pm , 
~~~~~~~~~~~~~~~
J_\pm (z) \Sh_\mp (0) \sim - \frac{2}{z} S_\pm
\\
\nonumber
&&
S_\pm (z) \Sh_\pm (0) \sim \pm \inv{z} J_\pm
\\ \nonumber
&&
S_+ (z) S_- (0) \sim  \frac{k}{z^2} + \inv{z} (H-J)
\\ \nonumber
&&
\Sh_+ (z) \Sh_- (0) \sim - \frac{k}{z^2} + \inv{z} (H+ J)
\end{eqnarray}

\def\Sb{\overline{S}}
\def\Shb{\overline{\Sh}}
\def\Hb{\overline{H}}

To better reveal the algebraic structure, we define new
couplings
\beq
g_v \CO^v + g_m \CO^m = g_+ \CO^+ + g_- \CO^- 
\label{newcoup}
\eeq
with 
\beq
g_\pm = g_v \pm g_m ,~~~~~~\CO^\pm = (\CO^v \pm \CO^m)/2 
\label{3.9}
\eeq 
The perturbing operators can then be written in terms of
currents in the following way:
\barray
\CO^+ &=& \Sh_+ \Shb_- - \Sh_- \Shb_+ + \inv{2}
\( J_+ \Jb_- + J_- \Jb_+ \) 
\nonumber
\\
\CO^- &=& J\Jb - H \Hb + S_+ \Sb_- - S_- \Sb_+ 
\label{3.11}
\\
\nonumber
\CO^a &=& (J-H)(\Jb - \Hb ) 
\earray 

Using the OPE's (\ref{3.8}), it is straightforward to 
compute the RG data from (\ref{2.5}, \ref{2.7}).  
One finds the non-zero values:
\barray
\label{3.12}
C^{a+}_+ &=& C^{+a}_+ = C^{++}_- = -4
\\
\nonumber
C^{++}_a &=& C^{+-}_+ = C^{-+}_+ = - C^{--}_a = -2 
\earray
and 
\barray
\label{3.13} 
\ct^{++}_+ &=& \ct^{+-}_a = \ct^{--}_a = - \ct^{-+}_+ = -2
\\
\nonumber
\ct^{+-}_- &=& \ct^{+a}_a = - \ct^{a+}_+ = -4 
\earray
The non-zero $D$'s are 
\beq
\label{3.14}
D^{a-}_a = D^{-a}_a = D^{--}_- = - D^{++}_+ = -1
\eeq
The matrix $D(g)$ is thus non-diagonal.  In a basis
$g=(g_+ , g_- , g_a ) $:
\beq
\label{3.15}
D(g) = \left( \matrix{ g_+ &0&0 \cr
                       0 & - g_- & -g_a \cr
                       0& 0&-g_- \cr }   
        \right)
\eeq
 
\def\gp{g_+}
\def\gm{g_-}
\def\ga{g_a}

After some algebra 
we find the \betaf functions:
\barray
\nonumber
\beta_{g_+} &=&\frac{
8g_+ \( g_+^2 ( 2g_a -g_- + 2) + 2 g_- (2-g_-) + 8 g_a \) }
{ (4-g_+^2 ) (2-g_-)^2 } 
\\
\beta_{g_-} &=& \frac{ 8 g_+^2 (2+g_-)^2 }
 {(4-g_+^2 )^2 } 
\label{3.16}
\\ \nonumber
\beta_{g_a} &=&  \frac{ 4 \( (g_+^2 - g_-^2)(16-g_+^2 g_-^2 )  
+ 4 \ga \gp^2 (2+\gm) (2-\gm)^2 \) }
{ (4-\gp^2)^2 (2-\gm)^2 } 
\earray

\subsection{Phases of the network model}

The \betaf functions (\ref{3.16}) have a precise  phase structure
that is not readily apparent from their form.  We studied the
phase structure by   
analyzing the behavior in the 
vicinity of the poles combined with some modest numerical work. 
Namely, the \betaf function differential equations 
 were solved numerically for 
$g(r)$ for a variety of points in each phase.  

\def\osp{osp(2|2)}

To begin describing the phase diagram we start with the solutions
to $C + \ct =0$.  There are two solutions: (i) $ \gp = - \gm , \ga = 0$
and (ii) $\gp = 0$.   The first solution corresponds to 
$g_v = 0, g_\pm = \pm g_m$ and the perturbation is thus $g_m \CO^m$
with  
\beq
\label{3.17}
\CO^m = \( -J\Jb + H\Hb - S_+ \Sb_- + S_- \Sb_+ 
+ \Sh_+ \Shb_- - \Sh_- \Shb_+ + \inv{2} 
  ( J_+ \Jb_- + J_- \Jb_+ )  \)  
\eeq
$\CO^m$ is built on the Casimir of $\osp$, so this corresponds 
to an $\osp$ symmetric manifold.  The \betaf function is 
\beq
\label{3.18}
\beta_{g_m} = \frac{-8 g_m^2}{(2+g_m)^2 } 
\eeq
When $g_a \neq 0$ this line is unstable so it does not play a 
significant role in the phase diagram.  However as we will see,
there is one regime that may be attracted to it.  

\def\gl{gl(1|1)}

For the  other solution $\gp = 0$ with $g_- , g_a$ arbitrary, the
only currents in the perturbation are $J, H, S_\pm$ which 
generate $\gl$ at level $k=1$.  The two operators $\CO^-$, 
$\CO^a$ correspond to the two independent Casimirs of 
$\gl$, reflecting the indecomposability of the adjoint representation. 
The \betaf functions reduce to 
\beq
\label{3.19}
\beta_{g_a} = - \frac{4 g_-^2}{(2-\gm)^2 }, ~~~~~~~~~~~\beta_{g_-} = 0
\eeq
This model was studied in \cite{GLL} in connection with the disordered
XY model and the localization problem of electrons randomly
hopping on a lattice
with $\pi$ flux per plaquette\cite{Hatsugai}.
  The coupling $g_+ = 0$ corresponds
to $g_v + g_m = 0$ which means that one of the couplings $g_{v,m}$ is
negative.  In \cite{GLL} such negative couplings arose naturally
in a different hermitian hamiltonian with twice as many degrees of
freedom.   In the model we are studying, as we will see, a large
regime of couplings, including initially all positive couplings, 
are attracted to this $\gl$ invariant manifold.  This is rather
unexpected:  starting from a hermitian hamiltonian with positive
variances $g$, in some regimes the couplings flow to negative values
which correspond to imaginary potentials and non-hermitian hamiltonians. 

There exists another $\osp$ invariant line which does not appear
as a solution to $C + \ct = 0$.  It arises due to the automorphism of
$\osp$:
\beq
\label{3.20}
J_\pm \to - J_\pm , ~~~~~~~ \Sh_\pm \to - \Sh_\pm 
\eeq
which does not change the algebra. 
Performing this automorphism on the left-moving currents only 
sends $\CO^m \to - \CO^v$.   Therefore the line $g_+ = g_-, g_a = 0$,
corresponding to $g_m = 0$, is also $\osp$ preserving with the
\betaf function:
\beq
\label{3.21}
\beta_{g_v} = \frac{8 g_v^2} { (2-g_v)^2  } 
 \eeq
For $g_v > 0 $, this is marginally relevant and is thus a massive
theory.  It's an integrable theory and the exact S-matrix was proposed
in \cite{Bassi}.  In the network model, one phase (called $O^\pm$
below) 
 appears to be
attracted to this line but with $g_v <0$, and thus flows to zero,
and is massless.  

\def\su{su(2)}

As in the $su(2)$ case the global features of the phase diagram
largely, but not completely,
 follow from the behavior near the poles $g_+, g_- = \pm 2$.  
Consider first the vicinity of the pole $(g_+, g_-) = (2,2)$.  
Letting $g_\pm = 2 + \ep_\pm$, one has 
\beq
\label{3.22}
\beta_{\ep_+} \approx \frac{-64 g_a}{\ep_+ \ep_-^2} , ~~~~~
\beta_{\ep_-} \approx \frac{32}{\ep_+^2} , ~~~~~~
\beta_{g_a} \approx \frac{ 16(1+ g_a)}{\ep_+^2}  - \inv{16 \ep_-^2 } 
\eeq
If $g_a > 0 $, then $\ep_+$ is attracted to zero, whereas $\ep_-$ 
always grows.  Here, $g_+$ flows to $2$ and $g_a, g_-$ flow to infinity
if $g_- >2$.
Since $g_+ / g_-$ flows to zero, we interpret this phase as flowing 
along the $\gl$ invariant line.   

The behavior around $(g_+, g_-) = (\pm 2, -2)$ is different.  Here one
finds:
\beq
\label{3.23}
 \beta_{\ep_+} \approx   \frac{-4 g_a}{\ep_+ } , ~~~~~
\beta_{\ep_-} \approx \frac{2\ep_-^2 }{\ep_+^2} , ~~~~~~
\beta_{g_a} \approx \frac{4\ep_-}{\ep_+}  
\eeq
 Again when $g_a > 0 $ one is attracted to $\ep_+ = 0$.  However
when $\ep_- = 0$, $\beta_{\ep_-} = 0$.

\def\ga0{g_{a0}}

A simplifying feature is that there are no poles in $g_a$. 
For each phase,
given  initial values for $g_\pm$, we find that for some
value $g_{a0}$ which depends non-universally on $g_\pm$, 
then $g_a > g_{a0}$ and $g_a < g_{a0}$ are distinct phases.
Examining $\beta_{g_+}$ near the poles $g_+ = \pm 2$, one finds 
\beq
\label{gaoo}
\ga0 \approx  (g_-^2 - 4)/8  ~~~~~~~~~~{\rm  near} ~ g_+ = \pm 2 
\eeq
For large $g_+$,  $\beta_{g_+} \approx -8g_+(2g_a - g_-)/g_-^2$, 
thus
\beq
\label{gaooo}
\ga0 \approx  g_-/2 ~~~~~~~~~{\rm for} ~~ g_+ \gg 1 
\eeq   
\def\den{\overline{\rho(E)}}

We list the regions with distinct behavior below. 
The phase diagrams are shown in
figures 2,3.     For each phase we determine the density of
states exponent.  To study the average density of states 
$\overline{\rho(E)}$ we shift $H \to H -E$ leading to a coupling
$E \Phi_E$ in the effective action with
\beq
\label{3.24}
\Phi_E = \psib_- \psi_+ + \psi_- \psib_+ + \betab_- \beta_+ 
            + \beta_- \betab_+ 
\eeq
The density of states is proportional to the one-point function of
$\Phi_E$:
\beq
\label{3.25} 
\den \propto \langle \Phi_E \rangle 
\eeq
Let $\Gamma_E$  denote the scaling dimension of the operator $\Phi_E$ 
in the IR.  Then since $E$ has scaling dimension $2-\Gamma_E$, 
\beq
\label{3.26}
\den \propto E^{\Gamma_E/(2-\Gamma_E)}  , ~~~~~~~{\rm as ~} E \to 0
\eeq

\bigskip
\noindent
$\underline{\bf g_a > \ga0 phases}$

\bigskip
\noindent
(i)  {\bf $dXY^\pm$ phases}. 
Here $g_+$ flows to $\pm 2$ after a finite scale transformation,
 and $g_-, g_a$ flow
to $+\infty$.  This is similar to the D region of anisotropic 
$su(2)$.    
The ratio $g_+ / g_- \to 0$,  so considering $g_+$ as effectively
zero, one interpretation of the flow is to the coset
$osp(2|2)_1 / gl(1|1)_1$.  As in the $su(2)$ case
this is not a true fixed point since the flow cannot be extended
to arbitrarily large length scales.   
The stress tensor for the $\gl_1$ current algebra conformal field theory
has a structure that parallels the structure of $\CO^- , \CO^a$.  Namely,
the stress tensor is an affine-Sugawara 
construction built on the sum of
the two independent Casimirs\cite{Roz}
\beq
\label{3.27}
T_{gl(1|1)_k}  = - \inv{2k} (J^2 - H^2 + S_+ S_- - S_- S_+ ) 
                   + \inv{2k^2} (J-H)^2  
\eeq
This stress tensor has $c=0$ so the coset $osp(2|2)_1 / gl(1|1)_1 $
also has $c=0$.   
The fermions $\psi_\pm$ have conformal dimension $\Delta = 1/2$
with respect to the $\gl$ thus the $\gl$ dimension of $\Phi_E$ 
is $1$.  This implies $\Gamma_E = 0$ and a constant density of 
states near $E=0$.  
Though the algebra $gl(1|1)$ is smaller than $osp(2|2)$,
the $gl(1|1)$ dimensions of fields are the same as for
$osp(2|2)_1$.  One can in fact  construct a level-1 representation
for both of these current algebras using the same number of
fields: two bosons and a complex fermionic scalar. (See e.g.
\cite{GLL}.)
Thus
$osp(2|2)_1 / gl(1|1)_1$ is empty and this would be  a massive phase.

\medskip

\noindent
(ii)  {\bf $O^\pm$ phases.}   
This phase is characterized by $(g_+, g_-)$ flowing to the pole
$(\pm 2, -2)$ while $g_a$ flows to zero.  
Unlike previous examples, here one is not attracted to the
isotropic lines $g_+ = \pm g_-$ before reaching the pole
since the line is unstable.   
The fate of the flow once it reaches the pole is again  delicate.  
Based on the example of $su(2)$ it seems most likely that this
flow spills into the $Q^\pm$ phases.   However here we do not have
the duality arguments to support this.  

\medskip
\noindent
(iii)  {\bf $Q^\pm$ phases.} 
In this region one flows to the poles at $g_+ = \pm 2$ after
a finite RG time but unlike the $dXY^\pm$ phases, numerical integration
indicates that  the other couplings
$g_-, g_a$ do not flow to infinity but rather to some finite
non-universal values as one approaches the pole.   Since none
of the couplings are flowing to infinity, we cannot interpret
this as a coset.        
This phase clearly requires further investigation.    
\def\rlim{ {\longrightarrow}   }

\begin{figure}[htb]
\hspace{18mm}
\includegraphics[width=12cm]{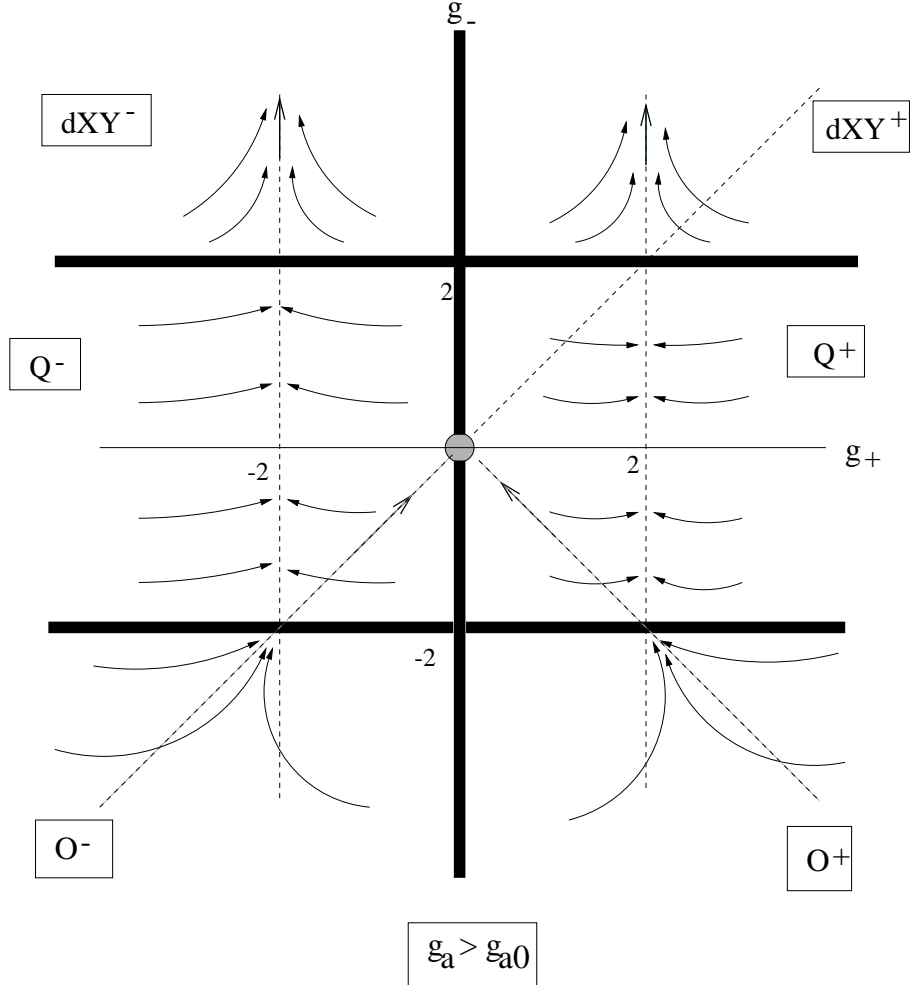}
\caption{Phase Diagram of the network model with $g_a > g_{a0}$.
 Phase boundaries
are solid heavy lines.
Shaded circles are fixed points. The cone opening to the right
is the physical regime of the network model.}
\label{Figure2}
\end{figure}

\bigskip\bigskip
\noindent 
$\underline{\bf g_a < g_{a0} phases}$

\medskip
\noindent
(i) {\bf $PSL_g$ phase}. 
Here $g_+ \to 0$, but $g_-$ flows to a fixed,
non-universal value.  $g_a$ flows very slowly to $-\infty$.  
In this phase $g_{a0} \approx 0$.  Unlike the flows to the poles,
in this case the flows can be run to arbitarily large length
scales.    For example
\beq
\label{flow1}
(g_+ , g_- , g_a ) = (1,1,-1)  \rlim  (0,  1.1508434.., -\infty)
\eeq
as $r\to \infty$. 
In this phase we are again on the $\gl$ invariant manifold but in
contrast to the $dXY$ phase there is a line of fixed points
corresponding to the value of $g_-$.  This is consistent with
the \betaf function (\ref{3.19}) since $\beta_{g_-} = 0$.  

\begin{figure}[htb]
\hspace{18mm}
\includegraphics[width=12cm]{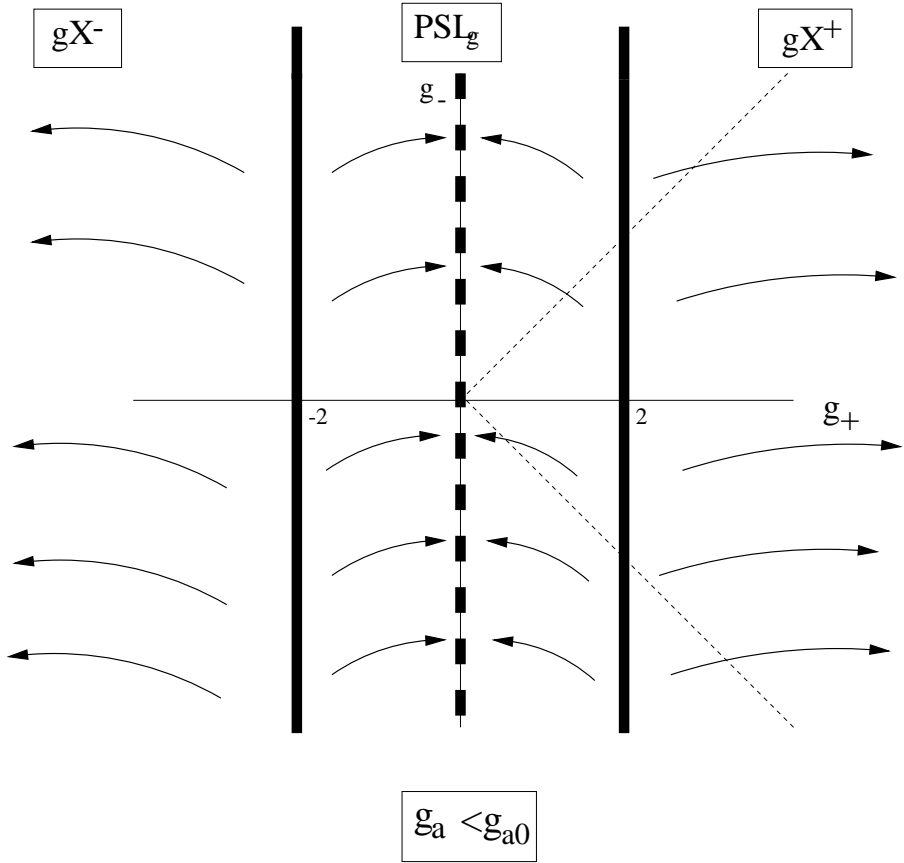}
\caption{Phase Diagram of the network model with $g_a < g_{a0}$. 
 Phase boundaries
are solid heavy lines.  
The heavy dashed line is a line of fixed points. 
}
\label{Figure3}
\end{figure}
The massive 
decoupled subalgebra $\ch$ is then generated by only $H-J$.  
However it 
 appears inconsistent to divide only by $H-J$.   The current
$H-J$ is not primary with respect to the stress tensor
$T = (J-H)^2/4$, i.e. $T(z) (H-J)(0)  \sim 0 $.  
A consistent conformal embedding is based on the subalgebra 
$\ch = u(1) \otimes u(1)$ with stress-tensor
\beq
\label{u1}
T_{\ch} = \inv{2} \( H^2 - J^2 \) 
\eeq
The fixed point is then:
\beq
\label{fixed2}
PSL_g ~ {\rm phase}:
   ~~~~~~ {\rm IR~ fixed~ point}=  ~~~
\frac{osp(2|2)_1} { u(1)\otimes u(1)}
\eeq
with $c=-2$.  This is a massless phase, since the above coset
is not empty.    In the formal limit $g_- \to \infty$ 
the above fixed point is the same as for the $dXY$ phases.   
As described in \cite{GLL},   dividing by $u(1)\otimes u(1)$ leaves
the free field theory of a complex {\it fermionic}  scalar,
with action
\beq
\label{c2}
S = \int  \d_\mu \chi^\dagger \d_\mu \chi
\eeq
This  theory is equivalent  
to  a $(\xi, \eta)$ fermionic system of conformal 
scaling dimension 
$(0,1)$\cite{FMS}.  It is perhaps the simplest logarithmic 
conformal field theory\cite{Gurarie}.  
 Interestingly the latter theory was used to describe
the dense phase of polymers by Saleur\cite{dense}.
This theory is also closely related
to the $PSL(1|1)$ sigma model introduced into the context of quantum Hall
transitions by Zirnbauer\cite{Zirnpsl}. 
The coupling $g_-$ naively corresponds to the exactly marginal 
direction $\delta S = g_- S$ where $S$ is given in 
(\ref{c2}), so that $g_-$ is similar to a radius of compactification
for a free boson.   However the modular invariant partition functions
at $c=-2$ do not have a continuous parameter, but rather are related
to Coulomb gas partition functions at certain discrete 
radii\cite{Kausch}. This suggests that as $g_a \to \infty$, $g_-$ 
can only take discrete values in order to lead to a modular
invariant partition function.  The spectrum of anomalous dimensions
is then the same as in \cite{dense}.  In particular   
the twist fields of this theory have fixed dimension 
$-1/4$ and  $3/4$\cite{dense}\cite{Gurarie}.

\medskip

\noindent
(ii) {\bf $gX^\pm$ phase. }    Here $g_+ \to \infty$,
$g_a \to -\infty$, and $g_-$ flows to a finite
non-universal constant   
as $r$ goes to $\infty$. 
For example:
\beq
\label{flow2}
(g_+ , g_-, g_a ) = (10,5,1) \rlim (\infty,  5.8829077.., -\infty) 
\eeq
Though $g_a $ flows to $-\infty$, the ratio $g_a/g_+$ flows to zero. 
Since  $g_- / g_+$ also flows to zero, we interpret this as 
$g_a =g_- = 0$.  

The coupling that goes to infinity, $g_+$, couples the currents
$\Sh_\pm , J_\pm $.  These do not form a closed subalgebra,
and for this reason this phase is more difficult to comprehend
and requires some speculation.  
 It could simply be a massive phase corresponding
to $osp(2|2)_1/osp(2|2)_1$
 since the above currents close on the whole of
$osp(2|2)$, as in the massive sine-Gordon phase 
of $su(2)$ (region E).  Alternatively let us suppose that only some  
of the currents are set to zero by $g_+ $ going to $\infty$.  Since 
the commutator of $J_\pm$ with $\Sh_\pm$ closes on $\gl$ currents
$S_\pm$, and the $\gl$ coupling $g_-$ is not flowing to infinity,
we cannot consistently set both $\Sh_\pm$ and $J_\pm$ to zero.  
To distinguish $\Sh$ and $J_\pm$ one may need some further $osp(2|2)$ 
symmetry breaking;  this could come from the fact that the operator
$\Phi_E$ breaks $osp(2|2)$.   
Two subalgebras of $\osp$ involving the above currents correspond to
another $\gl$ generated by $(\Sh_\pm , J, H)$ 
and $su(2)\otimes u(1)$ generated by $(J_\pm , J, H)$.  
Let us suppose the $\gl$ is set to zero by $g_+ \to \infty$.
As for the $dXY$ phase, the coset $osp(2|2)_1/gl(1|1)_1$ 
is empty and this would be a massive phase.  

Let us consider the other possibility.  
The currents $J, J_\pm$ generate an $su(2)$ at level $k=-1/2$.  To 
see this,  let  $J\to 2 J$, $J_\pm \to \pm 2 \sqrt{2} J^\pm$.
Then the new currents satisfy the OPE (\ref{4.1}) with 
$k= -1/2$.  The coset in this case is 
$osp(2|2)/su(2)_{-1/2}\otimes u(1)$.      
The central charge of $su(2)_k$ 
is $c=3k/(k+2)$ which gives $c=-1$ for $k=-1/2$.  Since $c=1$ for the 
$u(1)$, the  above coset has  $c=0$.   
The left-moving conformal dimension of primary fields of spin $j$ 
in $su(2)_k$ current algebra is 
$$\Delta^{(j)}_k = \frac{ j(j+1)}{k+2}$$
which equals $1/2, 4/3, 5/2, ...$ for $k=-1/2$.  The ghost fields
$\beta_\pm$ transform in the spin $1/2$ representation.  Thus in the 
IR the operator $\beta_+ \betab_-$ has dimension 
$\Delta^{osp(2|2)} - \Delta^{su(2)} = 0$.  Dividing by $u(1)
$ also leads to dimension zero for the fermion part of $\Phi_E$.  
Thus again $\Gamma_E =0$ and $\den $ is  a constant.  
Since the $su(2)_{-1/2}$ is built directly from the ghost
fields with the ghost fields in the spin $1/2$ representation,
the $osp(2|2)_1$ theory appears to be equivalent to 
$su(2)_{-1/2} \otimes u(1)$ if the only primary field 
of $su(2)_{-1/2}$ is $j=1/2$.   The coset 
$osp(2|2)_1/su(2)_{-1/2}\otimes u(1)$ is thus empty
and this is also a massive phase.  

\subsection{Physical regime of the network model} 

The physical regime of the network model is $g_v, g_m , g_a$ all positive. 
In the $(g_+ , g_-)$ plane this regime of couplings 
corresponds to the intersection of 
$g_+ > g_-$ and $g_+ > - g_-$, which is the
 $90^\circ$ cone symmetric about the 
$g_+$ axis. For reasons we do not yet understand, 
 couplings can flow in or out of this cone.  Let us
suppose that the initial couplings are in the cone, and furthermore
that initially $g_a > 0$.  The phases $dXY^+$, $Q^+$ 
 and $O^+$ are easily
accessible since $g_a$ is positive and can easily be chosen greater
than $g_{a0}$. 

Consider now the $g_a < g_{a0}$ phases.
For $PSL_g$, $g_{a0}$ is $(g_-^2 - 4)/8$ near $g_+ = 2$.  But
since $g_-^2 < 4$ in the cone, this phase appears inaccessible to
the network model.  If a negative $g_a$ turns out to be physically
sensible, perhaps for the reasons described in \cite{lecc}, 
then this phase can be realized.       
On the other hand, $g_a < (g_-^2 - 4)/8$ is easily satisfied with
a positive $g_a$ in the cone for the $gX^+$ phase.  An example is
equation (\ref{flow2}).  

In summary, the physical regime of the network model can flow to 
one of 4 different phases $dXY^+, O^+$,  $gX^+$, and $Q^+$.  
Initially
weak couplings flow to the phase
$Q^+$,  which, as stated above, cannot be interpreted with our
hypotheses since none of the couplings are  flowing to infinity.

In the above phases $dXY^+$ and $gX^+$, 
the density of states is not critical, 
i.e. $\den \propto E^0$ since $\Gamma_E = 0$.  This 
is in accordance with the expectation that the disordered
1-copy theory is not critical.  
The conventional wisdom is that one needs to study the 2-copy theory
and compute disorder averages of the product of retarded and
advanced Green functions. This can be done using the 
methods of this paper.  Since the 1-copy theory is contained in
the N-copy version,  one expects on physical grounds that the 
\betaf functions are the same for all $N$.  We checked that this
turns out to be the case and is a consequence of the zero super-dimension
of $osp(2N|2N)$.  
Though the \betaf functions are the same as we described for $N=1$,
the fixed points are different since the current algebras involved
are different, in particular, $osp(2|2)_1$ is replaced by
$osp(2N|2N)_1$.  This will be described in a separate publication. 

Though the localization length exponent $\nu$  is currently beyond our
understanding,  let us examine the possible exponents for the
dense polymer phase ($PSL_g$).   Tuning through the critical
point generally corresponds to a perturbation  
$$\delta S =  \epsilon \int d^2 x ~   \Phi_\epsilon$$
 with 
$\epsilon \approx 0$ for some operator $\Phi_\epsilon$.  
Let $\Gamma_\epsilon$ denote the scaling dimension of 
$\Phi_\epsilon$ at  the infra-red fixed point.
  The {\it mass} dimension of 
$\epsilon$ is then $2-\Gamma_\epsilon$.  As $\epsilon \to 0$, 
there is thus a diverging length scale  
\beq
\label{xi} 
\xi \propto  \epsilon^{-\nu} , ~~~~~~\nu = 1/(2-\Gamma_\epsilon)
\eeq
The  fields in the dense polymer theory  all have dimension
$n/16$ plus an integer.  In particular there is a field of
dimension $\Gamma = 25/16$,   corresponding the conformal dimension
$h = 25/32$,   which is a descendent ($1+$) of the field which is 
the dense polymer 1-leg operator ($h=-3/32$)  times  a twist
field ($h=-1/8$) from the  $\chi$-theory sector of another copy.
Note that this field does not exist in the 1-copy theory. 
Such a field leads to 
$\nu = 16/7$.  Within this scheme, 
this appears to be the closest one can  come to
the numerical value  $2.35 \pm .03$\cite{Huck}.  Unfortunately we have
no further arguments supporting the significance of this operator.

\section{Conclusion}

Under the hypotheses outlined above we have interpreted most of 
the RG flows based on  the all-orders \betaf functions 
proposed in \cite{GLM}.       
Much remains to be further clarified, in particular the 
flows
that are attracted to poles in the \betaf function after a finite
scale transformation need further investigation.           
In any case, this work  shows that disordered fermions in $2D$ at
strong coupling can have a  rich phase structure.  
In the physical regime of the network model
the important phases that we could understand as cosets 
appear to have  a constant
density of states, in accordance with the conventional
understanding.   
In addition, 
our analysis of anisotropic $su(2)$ could have implications
for Kosterlitz-Thouless transition physics at strong coupling. 

The only certainly massless phase we found for the network model
is the $c=-2$ conformal field theory $osp(2|2)_1/u(1)\otimes u(1)$,
which is known to correspond to {\it dense} polymers.  
There exists a classical percolation picture for the quantum Hall
transition\cite{Trugman}, and it is known that percolation
and dilute polymers are closely related $c=0$ conformal field 
theories\cite{dense}.  Our work seems to suggest that with strong disorder
the classical dilute polymer theory  flows to a dense polymer phase.

We considered the simplest case of one copy of Dirac fermion. 
In the theory of disorder,  for the same model one needs to 
study more copies  to compute averages of products of correlation
functions, and this generally has multifractal behavior.  
For $N$ copies 
the short distance unperturbed theory has $osp(2N|2N)_1$ 
current algebra symmetry.  The $N$-copy version of the 
network model, where one expects non-trivial exponents, 
will be studied in a forthcoming publication. 
The scheme described in this
paper can  lead  to
a classification of disordered critical points that parallels
the classification of sub-current-algebras $\ch_k$ of
$osp(2N|2N)_1$.  For this, the dictionary in \cite{diction}
is useful.    It would be interesting to compare this classification
with the classification based on sigma
models\cite{AlZirn}.   The latter classification is based on
discrete symmetries such as time-reversal, so it is not as
strong as our classification of the  actual critical points.  

It would be very interesting to perform more extensive numerical
simulations of the network model that vary the relative strengths of
the types of disorder and thereby see the phases predicted
in this paper.  For instance the $gX^+$  phase is characterized
by the randomness in the flux per plaquette and the tunneling
dominating over the randomness in the individual link phases.  
 
Though the models we  discussed are not integrable for general
anisotropic  
couplings,  under the 
RG they can  flow to isotropic  current-current
interactions and these are generally thought to be 
integrable\footnote{Note added in proof:  The $g\to 1/g$ duality
of the $su(2)$ model can be extended to the network model and
this leads to a resolution of the flows toward the poles. 
This will be reported on in \cite{BL2}.}

\section{Acknowledgments}

I especially wish to thank Denis Bernard for our 
collaboration on the spin Quantum Hall effect\cite{SQHE},  and 
Sathya Guruswamy and Andreas Ludwig for collaboration on
the disordered $XY$  model\cite{GLL}, both of which strongly influenced
this work.  I am also grateful to Phil Argyres, Zorawar Bassi, 
John Cardy, John Chalker, 
Vladimir Dotsenko, 
Bogomil Gerganov,  Jan\'e Kondev, Marco
Moriconi, and Tony Zee for discussions.    
I also want to thank the Centre de Recherche Math\'ematiques 
at the Universit\'e  de  Montr\'eal for holding a stimulating
workshop on related topics.  This work is in part supported by
the NSF.

\end{document}